\documentclass[oribibl]{llncs}

\usepackage{url}
\usepackage{amsmath}
\usepackage{balance}
\usepackage{natbib}
\usepackage{booktabs}
\usepackage{tabularx}
\usepackage{graphicx}
\usepackage{threeparttable}
\usepackage{natbib}
\bibpunct[, ]{(}{)}{;}{a}{}{,}

\usepackage[colorlinks=true,linkcolor=cyan,
  citecolor=cyan,urlcolor=cyan]{hyperref}

\newcommand{\change}[1]{\textcolor{black}{#1}}

\begin{document}

\title{Reassessing Measures for Press Freedom}

\author{Jukka Ruohonen\orcidID{0000-0001-5147-3084} \email{juanruo@utu.fi}}
\institute{University of Turku, Turku, Finland}

\maketitle
\thispagestyle{plain}

\begin{abstract}
  There has been \change{an increasing} interest in press freedom in the face of
  various global scandals, transformation of media, technological change,
  obstacles to deliberative democracy, and other factors. Press freedom is
  frequently used also as an explanatory factor in comparative empirical
  research. However, validations of existing measurement instruments on press
  freedom have been far and few between. Given these points, this paper
  evaluates eight cross-country instruments on press freedom in $146$ countries
  between 2001 and 2020, replicating an earlier study with a comparable research
  setup. The methodology is based on principal component analysis and
  multi-level regression modeling. According to the results, the construct
  (convergence) validity of the instruments is good; they all measure the same
  underlying semi-narrow definition for press freedom elaborated in the
  paper. In addition, any of the indices seems suitable to be used
  interchangeability in empirical research. Limitations and future research
  directions are further discussed.
  \vspace{5pt}
\end{abstract}

\section{Introduction}

Validation of empirical measures has a long history in sciences and scholarship,
from social sciences to engineering. Whether it is democracy~\citep{Vaccaro21},
corporatism and wage-setting~\citep{Kenworthy01}, or properties of software
systems~\citep{Kitchenham95}, a measure must be validated, either theoretically
or empirically, before it can be further used in empirical research and
data-based applications. Freedom of the press is not an exception. Thus, this
paper presents an empirical validation of eight cross-country indices for press
freedom in $146$ countries between 2001 and 2020.

Press freedom has a long research lineage in comparative communication and media
studies, \change{although, at the same time, empirical research has also long
  been constrained by the lack of valid empirical data.} The probably earliest
large-scale comparative study was conducted by \citet{Nixon60} who compared
press freedom in sixty-one countries in relation to other macro-level indices,
such as adult literacy levels and per capita national incomes. Similar studies
have been conducted ever since; correlations with press freedom have been sought
from corruption~\citep{Dutta16}, stock prices~\citep{Masrorkhah17}, and
terrorism~\citep{Hoffman13}, to only name three examples. Despite this empirical
research activity, two things stand out from the literature. First, theoretical,
legal, and philosophical work on freedom of the press has \change{arguably been}
neglected for a long time---press freedom was considered more or less a settled
issue in the disciplines mentioned. In fact, the amount of papers published on
press freedom in these disciplines has long lost to human rights, other
freedoms, and civil liberties~\citep{Lamer16}. Second, the empirical work in
other disciplines has been disadvantaged by a visible lack of validation
research. Correlations have been sought, but validity of the measures used has
remained open to criticism. In fact, according to a fairly comprehensive
literature search, there \change{are only two comparable validation
  studies}---the \change{papers} of \citet{Becker07} \change{and
  \citet{WhittenWoodring17}}. \change{In general, both papers found similarities
  across indices} provided by different organizations. As such, \change{these
  findings are} important contributions to the empirical comparative research on
press freedom. \change{These are worth replicating. As has been frequently
  argued, there is a continuous need for replication studies throughout all
  scientific and scholarly disciplines. Media and communication studies are not
  an exception~\citep{Benoit08}.} But there is more, much more to motivate the
study.

The discussion about press freedom started to change soon after \citet{Becker07}
had published their study. While the Satanic Verses was published already in
1988 and $9/11$ happened in 2001, it was in the 2010s when things started to
escalate; WikiLeaks and Chelsea Manning in 2010, Edward Snowden in 2013, Charlie
Hebdo in 2015, and so forth. These events and the global controversies that
followed spiced up also the academic discussion. As asked by \citet{Lee15}, what
does the asymmetry between satire and physical violence tell about press
freedom, given that satire is also asymmetrical, offering a weapon for mocking
the important and powerful?  As continued by \citet{Cushman16}, why was the
killing of journalists, editors, cartoonists, and policemen partially met with a
seemingly contradictory impulse to restrain the freedom of expression? Then
there are the questions related to national security, which, according to many
recent authors, pose a real danger to press freedom particularly through
self-censorship~\citep{Burman17}. \change{Here, the so-called chilling effects are
  well-known~\citep{Penney16, Stoycheff19}.} National security together with
leaks of confidential information have also spurred the legal and philosophical
research~\citep{Ogbebor20}. Although it is, in the present work, unnecessary to
delve into these topics in detail, these still establish a solid further
motivation for the empirical validation pursued.

\begin{figure}[th!b]
\centering
\includegraphics[width=\linewidth, height=5cm]{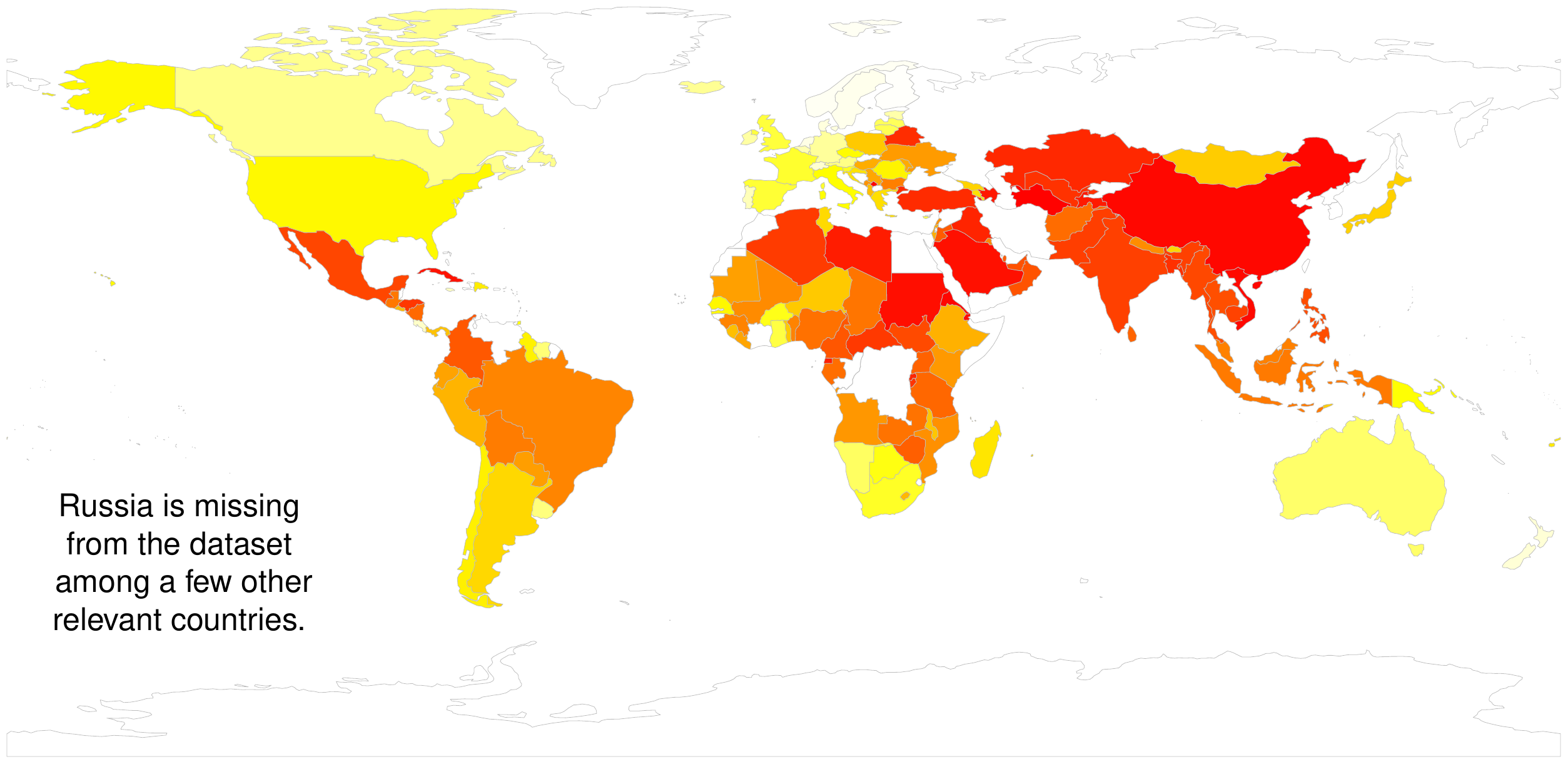}
\caption{Press Freedom Across the Globe in 2020 \citep{RSF21a}}
\label{fig: map}
\end{figure}

But there is more. The weakening of democracy and increasing of authoritarianism
\change{have} decreased freedom of the press worldwide~\citep{Lipman10,
  Simon06}. As can be concluded from Fig.~\ref{fig: map}, where a darker color
implies less freedom, the situation has remained constrained in much of Asia,
Middle East, and Africa. But in Europe, too, curtailment of press freedom has
occurred in Belarus, Ukraine, Hungary, Poland, and the Balkans. Given that the
paper's purpose is empirical validation of press freedom measures, a grain of
salt should be taken with the figure, but still only a grain.

Even in countries with a well-established and heralded press freedom doctrine,
such as the United States, threats and attacks against journalists increased in
the late 2010s~\citep{Jamil20}. Populism and extremism are partially to blame
for the increased hostility toward press in many Western countries. Another
partial explanation stems from the rise of social media and large platform
companies, engagement and emotionality, and associated factors related to
technology, which have made harassment of journalists easier and supposedly also
contributed to self-censorship. \change{Also the enduring digital divide has
  constrained the emergence of new online media outlets in developing
  countries~\citep{Ruohonen21ICSET}.} While journalists often consider
themselves as paragons of objectivity and the freedom of expression, also
economic pressures and political biases, which are often related, have
contributed to market-based censorship, ethical lapses, and even criminal
conduct~(\citealt{Ogbebor20}; \citealt{Petley12}). National cultures offer a
further partial explanation. According to interviews, journalists in many
countries tend to share the same universal principles about press freedom, but
these are often domesticated with specific national interpretations and
worldviews~\citep{Fahmy07, Voltmer14}. Somewhat similar findings have been made
in quantitative research; media freedom correlates with measures on national and
indigenous cultures, thus undermining the assertion of globalization as a force
devouring everything in its path~\citep{Dutta13}. Analogously, support for press
freedom by citizens tends to vary across countries as well as socioeconomic
factors within countries~\citep{Nisbet11}. Thus, all in all, there is hardly a
single explanation for the decline or stagnation of press freedom in the world.

The preceding contextual points justify the empirical validation of the eight
press freedom indices. But there is also a scientific justification; the
progress of science and the cumulation of empirical evidence necessitate
replications of existing results. To this end, \citet{Becker07} concluded that
organizations reporting on press freedom largely agreed with each other in terms
of their empirical datasets. When evaluated against each other, the eight
indices can be therefore expected to be valid. If this expectation does not
hold, a shadow is cast on the empirical research branch operating with press
freedom indices. For evaluating the expectation of validity, the following
section outlines the background in more detail. The materials and methods are
subsequently elaborated. Results and conclusions follow in their consecutive
sections.

\section{Background}

A~measure can be defined as a value of an attribute outputted by a given
measurement instrument~\citep{Kitchenham95}. A thermometer measures
temperatures; a counter measures the killing of journalists. Then, comparative
validation can be understood to involve multiple measures, attributes, or
instruments. The last case is likely the most common; whenever there are
multiple instruments supposedly measuring the same abstraction, such as a
theoretical concept, both theoretical and empirical assessments are
needed. Typically, the validation assessments are about construct validity; the
goal is to assert that a set of measures indeed measure the same abstraction of
interest. In terms of theoretical validation, in some settings, it is possible
to derive rigorous theoretical criteria for the validity of measures. For
instance, some measurements must be monotonic and non-negative with clearly
defined maxima and null values, and, thus, any new instrument proposed should
satisfy these theoretical properties~\citep{Briand98}. If no journalists have
been killed in a given country, the counter should consistently output a value
zero; the counter should not exceed the amount of journalists in the country;
and so forth. Although theoretical validations are commonly done when a new
instrument is developed, empirical validations have been more common in applied
research using already developed instruments and their measurements. Empirical
validation is also the goal of the present work.

A few sentences must be still written about the theoretical footings behind the
concept of press freedom. One way to think about the concept is to consider a
hierarchy or an inheritance tree of freedoms. Another way is to think about
different supersets and subsets of freedoms~\citep{Ruohonen21MIND}. Either way,
the analytical reasoning leads to the same conclusion: The freedom of thought
can be seen as the parent concept inherited by other freedoms, among these the
freedom of expression, which, in turn, is inherited by the freedom of
speech. Then, according to classical theorists, including John Stuart Mill, in
particular, freedom of the press is an extension of the freedom of speech; the
press has a right to self-expression like individuals---its voice is just louder
than the voice of any given individual alone, and, therefore, defending press
freedom is tantamount to defending the freedom of
speech~\citep{Mokrosinska20}. That said, there is a debate whether press freedom
and the freedom of speech can and should be equated together---a question that
becomes more interesting when a press is considered to be a political and social
institution instead of being merely a human rights
abstraction~\citep{Lamer16}. The rise of authoritarianism has further
complicated the equation of the two concepts; there may be freedoms for
expression and speech but no press freedom in some
countries~\citep{Lipman10}. Moreover, marrying the two freedom concepts leads to
an overly broad definition for any measurement purposes.

A narrower definition considers freedom of the press via independence. To be
free, a press needs autonomy from any political interference, including explicit
governmental censorship or implicit self-censorship, legal repercussions such as
licenses for publishing or liabilities from publishing, harassment or worse, and
related oppressive measures taken by a government, a political
establishment~\citep{Kenny19}, or fellow citizens. In a democracy both the
government and the press are for the people, but if one has to choose between a
government without a press and a press without a government, many would follow
Thomas Jefferson and ``not hesitate a moment to prefer the
latter''~\citep[quoted from][]{Chu87}. This assertion together with the loudness
of the voice given for people make it understandable why press freedom is
conventionally seen as essential for democracy. Without a loud ``aggregated''
voice of people through a press, which has traditionally represented the public
sphere in a democracy, there are only limited opportunities for most individuals
to participate in democratic deliberation and policy-making~(for the background
see \citealt[pp.~53--71]{Ogbebor20}). Although the Internet and social media
have somewhat changed the situation throughout the world, the basic premise
still arguably holds; a free press is necessary for any democracy. The point
serves also to make a further framing to the paper's scope: Unlike in some
previous studies~\citep{Kenny19}, the empirical validation only considers
``traditional'' measures for freedom of the press, omitting anything
``digital'', including governmental censorship of online-only newspapers and
outlets, content takedowns, Internet shutdowns, and related threats to press
freedom---and other freedoms.

Construct validity, too, is a parent concept inherited by other validity
concepts, such as convergence validity; the degree to which two or more
theoretically related constructs are empirically
related~\citep{Schimmack10}. This child concept provides the underlying
hypothesis for the present work: Because all of the instruments used have been
designed to measure the same semi-narrow definition for press freedom, these
should be also empirically associated with each other. Convergence validity
further establishes a practical rationale related to interchangeability: If the
values from the indices are similar, any index will do in empirical
applications~\citep{Vaccaro21}. \change{This provides a further contribution of
  the paper; the goal is not limited to convergence validity alone.} In order to
provide a sensible but still prototypical empirical application, the press
freedom instruments considered are used in a small regression model explaining
the working of deliberative democracy. As said, deliberation is closely related
to freedom of the press, and, therefore, the motivation is clear; the more there
is freedom for a press, the more there is room for free public
deliberation---even though press freedom is arguably only a necessary but not a
sufficient condition for deliberative democracy. When considering the following
definition, \change{many} of the threats to press freedom discussed---from
harassment to emotional appeals---\change{can be seen also as} threats to
deliberative democracy:

\begin{quote}
``\textit{The deliberative principle of democracy focuses on the process by
    which decisions are reached in a polity. A deliberative process is one in
    which public reasoning focused on the common good motivates political
    decisions---as contrasted with emotional appeals, solidary attachments,
    parochial interests, or coercion. According to this principle, democracy
    requires more than an aggregation of existing preferences. There should also
    be respectful dialogue at all levels---from preference formation to final
    decision---among informed and competent participants who are open to
    persuasion.}'' \citep[pp.~44--45]{VDem21b}
\end{quote}

Finally, reliability is a necessary prerequisite for all validity constructs. A
measurement instrument is reliable when it always---no matter the environment or
the user of the instrument---outputs the same value for the same measurement
object, all other things being constant; such an instrument is free from random
error~\citep{Hamann13}. Although reliability must be taken for granted for the
present empirical validation, it is necessary to say a few words about the
underlying research paradigm behind the instruments; these are cross-country
indices designed to be used in comparative research. For a given instrument,
each country is assigned a value in a given year by an expert or a group of
experts. Thus, a degree of random error is certainly present, as is the case in
social sciences almost always. But, in addition, a degree of politics is also
involved; even with the narrow definition, many of the definition's
tenets---such as harassment---are highly politicized, and some of the
tenets---such as lethal violence against journalists---are bleak media
spectacles in domestic and international politics. These points not only point
toward potential reliability issues but also restate the problems with the
narrow definition. According to critical arguments, freedom of the press
involves also such as things as plurality of viewpoints expressed in media,
meditation between societal conflicts, enlightenment, access to media content by
people, and even nation-building~\change{\citep{Becker07, Ruohonen21ICSET,
    Walters20}}. To account for these arguments, the indices used follow a
semi-narrow definition, augmenting governmental threats such as censorship with
a few aspects on plurality, criticality, and related broader media
concepts. While keeping this \change{point} in mind, \change{debates continue
  about definitions for press freedom}, as will be later briefly discussed in
the concluding section.

\section{Data and Methodology}

\subsection{Datasets}

Four datasets are used for the comparison. These can be summarized as follows:
\begin{enumerate}
\item{The first dataset covers the world press freedom index compiled by
  \textit{Reporters sans fronti\`eres} (RSF), a non-governmental organization
  specialized in media freedom and related freedoms. Due to online data
  availability issues, however, only data for the year 2021 was obtained
  directly from the RSF. Although also the remaining data (for the years between
  2001 and 2019) comes from the RSF, it was downloaded from the open online data
  portal provided by the \citet{WorldBank21a}. Irrespective of the download
  location, the RSF data is based on expert surveys delivered in twenty
  languages to professional journalists in nearly two hundred countries. The
  questionnaire used for the surveys is fairly extensive. \change{The 2021
    questionnaire covered} seven categories: legislation, infrastructure,
  independence (including economic and political independence), pluralism of
  opinions expressed in media, self-censorship, transparency, and---last but not
  least---abuse of journalists and media employees, including imprisonment and
  lethal violence~\citep{RSF21b}. Data for the last category is further
  monitored by a separate expert network. \change{Each index is measured with a
    scale from 0 to 100, higher values indicating worse conditions. A composite
    index constructed from the individual indices is used for the present
    work. A major limitation with the RSF's data is that the organization has
    continuously made alterations to its methodology. In 2021 the organization
    noted that a similar methodology was used between 2013 and
    2021~\citep{RSF21b}, but the methodology was again changed in
    2022~\citep{RSF22}. Thus, it remains unclear how press freedom was measured
    prior to 2013. All traces have disappeared from the organization's website
    about past datasets and their methodological details.}}
\item{The second dataset originates from the \citet{FreedomHouse23a}, another
  non-governmental organization specialized in democracy, political freedom, and
  human rights. The index used (D1) measures the freedom and independence of the
  press by considering the absence of lawsuits, intimidation, and physical
  attacks, \change{censorship and self-censorship, surveillance, independence
    from the government, financial pressure, ownership structure of media,
    editorial freedom, pluralism of viewpoints, gender issues, and many other
    factors~\citep[p.~1445]{FreedomHouse20}. Analogously to RSF, the data is
    based on expert evaluations involving in-house analysts, consultants,
    academics, and others. Unlike with RSF, however, Freedom House's methodology
    has remained fairly consistent over the years. While small alterations have
    been made, the evaluation criteria was highly similar in 2013 than it was in
    2022 \citep[cf.][pp.~851--852]{FreedomHouse13}. The index used ranges from 0
    to 7, lower values indicating worse conditions.}}
\item{The third dataset is based on the celebrated varieties of democracy
  (\text{V-Dem}) project \citep{VDem21a}. Currently, the project provides
  presumably both the most robust and the most comprehensive data on various
  measures related democracy, among these indices on civil liberties and various
  other concepts. Like with the RSF and the Freedom House, the data from
  \text{V-Dem} is based on expert assessments, but unlike in the two other
  projects, the assessments are done by independent academics and their
  staff. Six individual indices are used: (1)~governmental \textit{censorship}
  of media, (2)~\textit{harassment} of journalists, (3)~\textit{self-censorship}
  by media, (4) media \textit{bias}, (5) plurality of \textit{perspectives}
  presented in print and broadcast media, and (6) \textit{critical} journalism
  in print and broadcast media~\citep[p.~37]{VDem21b}. \change{The scale of the
    variables is such that increasing values imply more freedom.} All six
  indices have been used in previous research~\citep{Kenny19}. When compared to
  the other two data sources, the availability of unaggregated indices is a
  clear advantage of the \text{V-Dem} project. It is also worth noting that
  \text{V-Dem} provides separate indices for the ``freedom of discussion'' for
  men and women, but these were omitted on the grounds of preferring the
  semi-narrow definition for press freedom outlined in the previous section. The
  same point applies to other related indices on civil liberties. \change{In
    addition, the variables were chosen with the interchangeability assessment
    in mind. In other words, censorship, harassment, self-censorship, media
    bias, and perspectives in media are all relevant for deliberative
    democracy.}}
\item{The fourth dataset contains secondary data for the interchangeability
  assessment. The dependent variable for the assessment is a deliberative
  democracy index from the \text{V-Dem} dataset. It is a composite index
  constructed from individual indices, which take also electoral democracy into
  account (for details see \citealt[pp.~44-45]{VDem21b}). Given that the
  intention is not to build a full model for explaining deliberative democracy,
  Gross Domestic Product (GDP) per capita \citep{WorldBank21b} is used as the
  single control variable.}
\end{enumerate}

All datasets have been widely used in existing research. Although the Freedom
House has long been the \textit{de facto} data source for indices on press
freedom and civil liberties in general \citep{Dutta13, Hoffman13, Nam12,
  Nisbet11}, more recently the \text{V-Dem} project has caught more
attention---if not even enthusiasm. Data from \text{V-Dem} has been also
previously validated in a comparable research setup~\citep{Vaccaro21}. It is
furthermore worth noting that data from the RSF has previously been mixed with
data from the Freedom House~\citep{Sandoval16}. \change{The RSF and Freedom House datasets were evaluated also by \citet{Becker07}.} But despite widespread use in
research as well as in media---or perhaps because of this popularity, criticism
of the indices has also been common.

On occasion, for instance, the civil liberties data from the Freedom House has
been accused of subjectivity and Western biases \citep{OkuliczKozaryn14}. To
some extent, however, such criticism is lightweight---if not frivolous; the
question about Western tenets behind press freedom has been debated for
decades---if not centuries~\citep{Chu87}, and the elementary methodological
criticism can be generalized practically to all cross-country comparative
research. Nevertheless, the more nuanced academic criticism should still be
acknowledged~(see, in particular, \citealt{Crasnow20}). To defend against the
criticism and the choice of the datasets, the goal of empirical validation can
be recalled. In other words, the purpose is not to evaluate the theoretical and
methodological groundwork behind the indices, but to validate these against each
other. If the values from the indices are similar, the indices from the three
datasets can be asserted to be valid \textit{vis-\`a-vis} each other---even when
all indices and their measurements would be theoretically or methodologically
invalid for measuring the same semi-narrow definition of press freedom.

\subsection{Processing}

The datasets were merged by including only those countries that were present in
all datasets. Even after this merging, which reduced the amount of data
substantially particularly from the \text{V-Dem} dataset, two practical problems
remained.

The first problem was missing values. The RSF dataset misses values for all
countries for the years 2010, 2011, and 2020. There are also some additional
missing values for some countries. Likewise, the secondary data for the GDP per
capita misses the year 2020 for all countries. To address these issues, the
missing values were interpolated with cubic splines using a well-known software
implementation~\citep{akima}. Following existing
research~\citep{Ruohonen17ELMA}, a few negative values from the interpolation
were truncated to zero, given that negative values are not possible for the
indices in question. \change{Due to the interpolation, some of the values may
  exceed the original upper limits of the variables. This point applies
  particularly to the RSF dataset.}

The second problem is related. After interpolation, the RSF and \text{V-Dem}
dataset contain full annual records for $146$ countries in a period between 2001
and 2020. But the Freedom House dataset is still limited only to a period
between 2013 and 2020 for the $146$ countries. As the amount of missing values
and their time period prevent reasonable interpolation, the evaluation involving
the Freedom House dataset is done separately by deleting the corresponding
values from the two other datasets. Excluding the secondary data for the
interchangeability evaluation, the Freedom House evaluation deals with $n =
1168$ observations, whereas the evaluation with the two full-length datasets
operates with $n = 2920$ observations.

\subsection{Methods}

The overall methodological research designs follows a conventional state-centric
approach in comparative communication research and comparative research in
general; countries are the units of analysis and the evaluative comparison is
carried horizontally across countries~\citep{Ebbinghaus98, Livingstone03}. This
approach also leads to a conventional problem; the lack of statistical variance
is a notable concern with all datasets. As such, the problem may only be
statistical; the lack of variance can be interpreted to imply that press freedom
has mostly remained stable in many of the countries observed. In any case, to
illustrate the problem, Table~\ref{tab: variance} shows the number of countries
for which the measurements attain a standard deviation of zero, meaning that
only a single unique value is present. As can be seen, the situation is
particularly problematic for the Freedom House dataset within which as much as
56\% of the countries show no variation in press freedom between 2013 and
2020. With respect to the \text{V-Dem} dataset, the lack of variance is more
common in a period between 2001 and 2021 than in a period covering the 2010s. If
variance alone would be taken as a sole evaluation criterion, the RSF dataset
would be preferable.
\begin{table}[th!b]
\centering
\caption{Number of Countries with Zero Standard Deviation}
\label{tab: variance}
\begin{tabular}{lrrrr}
\toprule
& 2001 -- 2010 & 2011 -- 2020 & 2013 -- 2020 & 2001 -- 2020 \\
\hline
RSF & 0 & 0 & 0 & 0 \\
Freedom House & 0 & 0 & 82 & 0 \\
Censorship (V-Dem) & 40 & 10 & 19 & 7 \\
Harassment (V-Dem) & 53 & 6 & 13 & 6 \\
Self-censorship (V-Dem) & 54 & 12 & 21 & 6 \\
Media bias (V-Dem) & 37 & 12 & 17 & 7 \\
Perspectives (V-Dem) & 74 & 24 & 32 & 20 \\
Critical journalism (V-Dem) & 67 & 28 & 32 & 23 \\
\bottomrule
\end{tabular}
\end{table}

Also the evaluation is carried with conventional statistical methods. Following
a recent comparative evaluation~\citep{Vaccaro21}, observing the empirical
distributions is a good way to start the statistical analysis. Then:
Correlations are typically seen as sufficient for evaluating convergence
validity~\citep{Schimmack10}, but dimension reduction techniques, such as
principal component analysis (PCA), are equally suitable for the
task. \change{As PCA evaluates linear combinations across many variables, it is
  ideally suited for the assessment. When compared to correlation analysis, the
  benefit from PCA is that all variables can be assessed in a single group.}
The evaluation is thus carried out with PCA separately across the $146$
countries and across the eight years. The share of variance explained by the
first principal component (PC) is used to report the results; if the indices
converge to the same theoretical abstraction, a large share should be explained
by the first PC. Unfortunately, sample correlation matrices cannot be used due
to the lack of variance in the cross-country computation. Although being
sensitive to outliers and different scales of the indices, covariance matrix is
therefore used instead; a lack of variance translates into zero covariance
between a given two indices. Sample correlation matrix is still used as the
input for the computation across years.

Finally, the interchangeability test is computed with a conventional multi-level
regression estimator, using a well-known software
implementation~\citep{lme4}. As the intent is to only evaluate
interchangeability, there are no particular reasons to consider a more elaborate
or robust regression estimator. Nor is there a strong mandate to focus on
different model specifications; the models estimated are so-called random
effects models with which random intercepts are specified for the countries and
years. In the same vein: The GDP per capita is the only control variable
included in the models. Recalling the lack of data in the Freedom House data,
the models are estimated for the whole period between 2001 and 2020 and a
shorter period between 2013 and 2021. To evaluate the regression models, the
regression coefficients and their signs are the primary
interest. \citeauthor{Schwarz78}'s \citeyearpar{Schwarz78} classical Bayesian
Information Criterion (BIC) is used to briefly check statistical performance
across the models estimated.

\section{Results}

\begin{figure}[th!b]
\centering
\includegraphics[width=\linewidth, height=10cm]{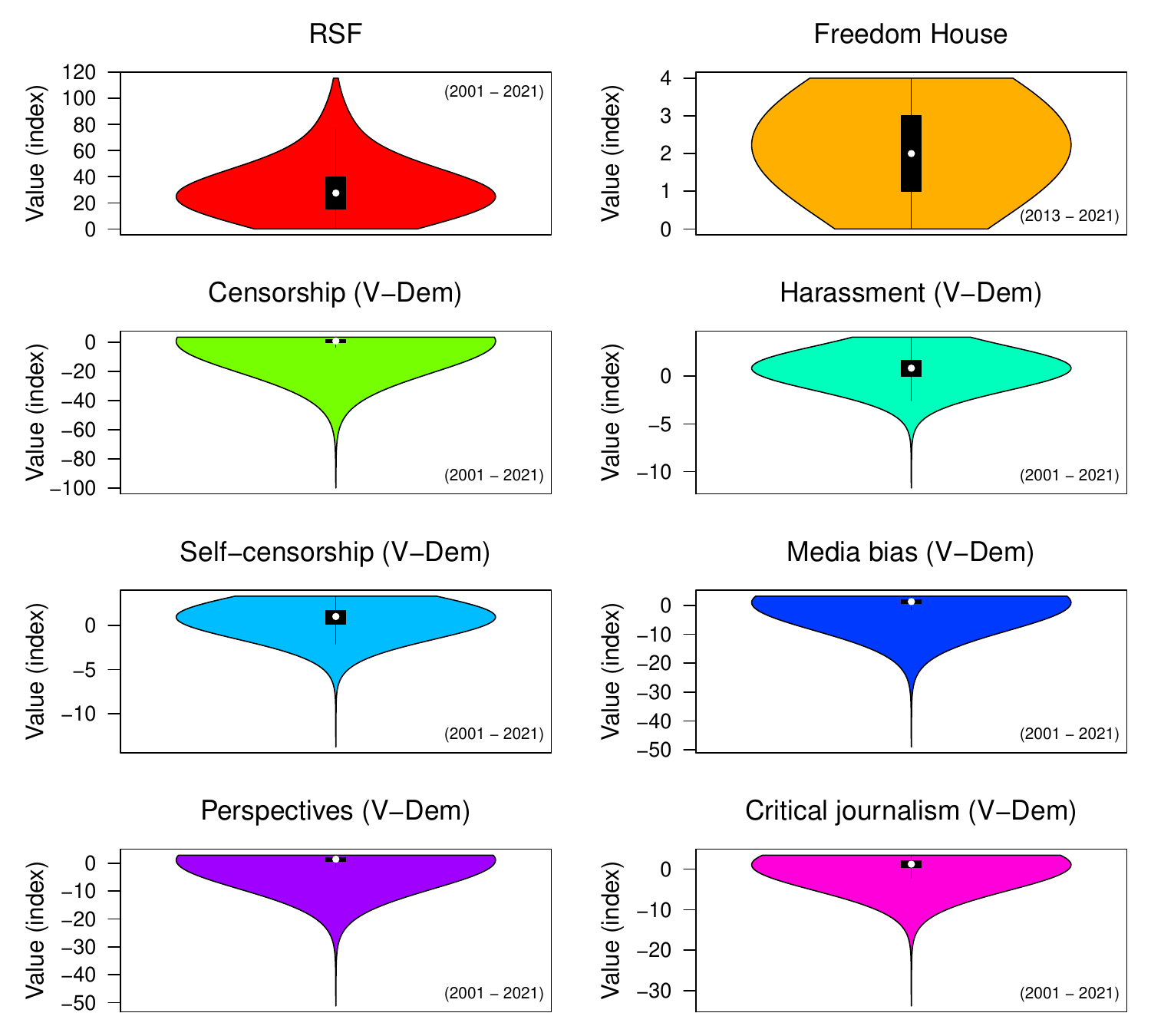}
\caption{Distributions of the Values from the Eight Indices}
\label{fig: violins}
\end{figure}

Empirical distributions of the eight indices across all countries and years are
illustrated in Fig.~\ref{fig: violins} by using so-called violin
plots~\citep{vioplot}. The shapes of the violins visualize the shapes of the
empirical probability distributions; the boxes represent the interquartile
ranges like in boxplots; and the dots within the boxes the medians. Like in
Fig.~\ref{fig: map}, higher values indicate less freedom in the RSF dataset; the
reverse applies to the other seven indices. Given these clarifications, it can
be concluded that only three indices show something remotely resembling a normal
distribution; the Freedom House index and the harassment and self-censorship
indices from the \text{V-Dem} dataset. All others violins display a long-tailed
distribution; there are many countries with a free press, some countries with a
moderately free press, and a few countries with a severely constrained
press. These observations align with Fig.~\ref{fig: map}.

\begin{figure}[p!]
%
\centering
\includegraphics[width=\linewidth, height=8cm]{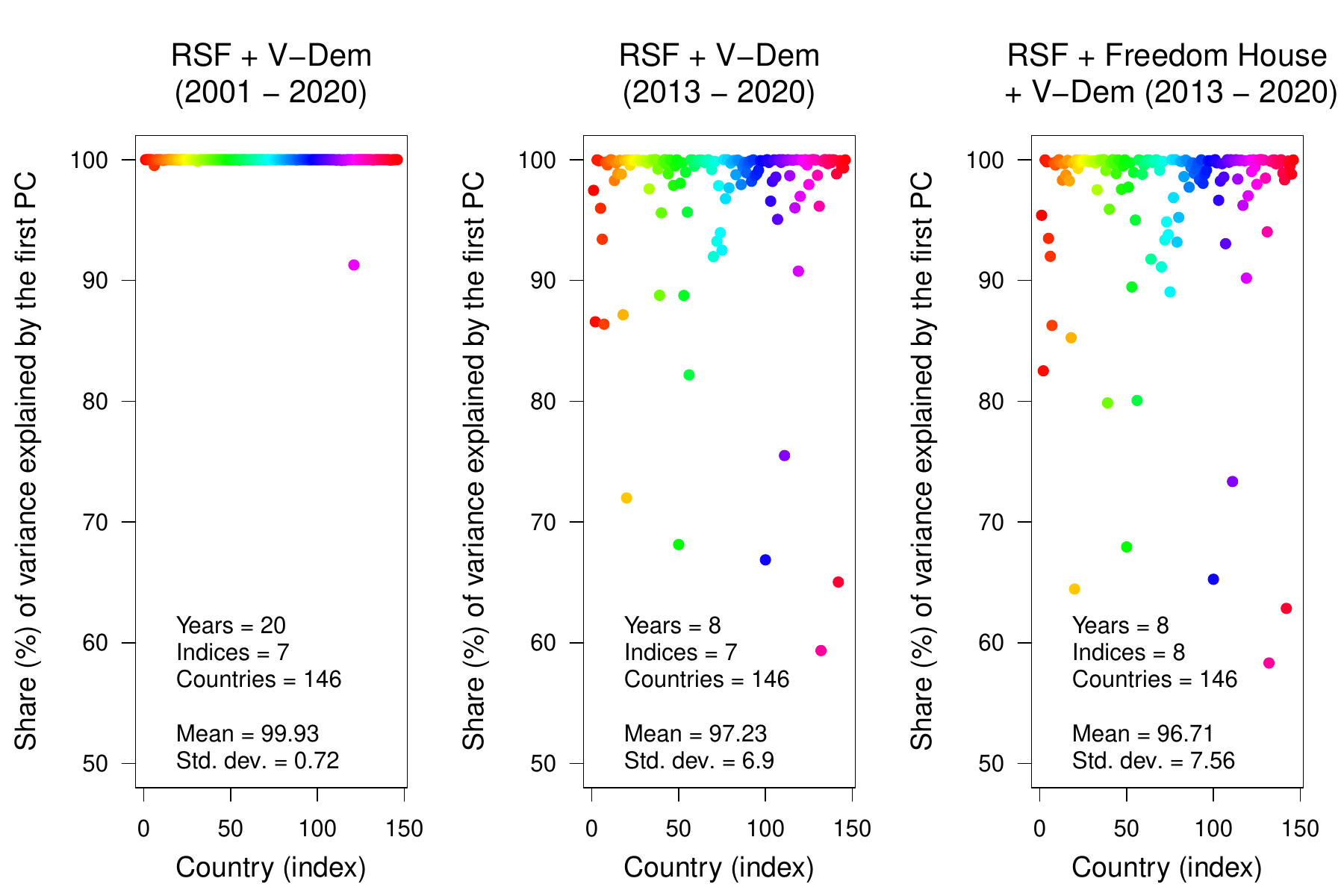}
\caption{Principal Component Analyses Across Countries (covariance matrices)}
\label{fig: pca countries}
%
\vspace{30pt}
%
\centering
\includegraphics[width=\linewidth, height=8cm]{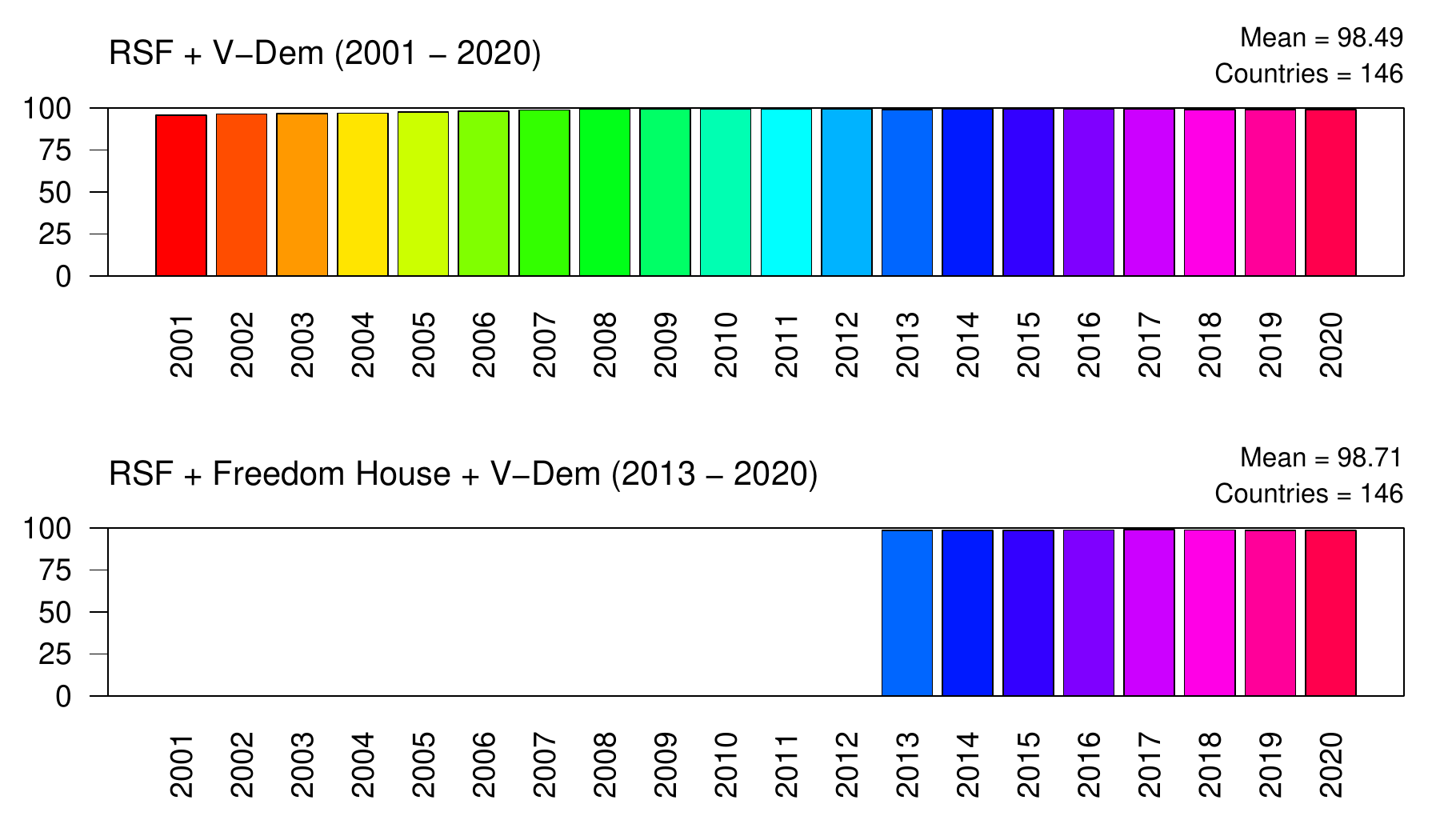}
\caption{Principal Component Analyses Across Years (correlation matrices)}
\label{fig: pca years}
%
\end{figure}

Turning to the convergence validity evaluation, the PCA computation across
countries is illustrated in Fig.~\ref{fig: pca countries}. To read the figure
from left to right, the first plot shows the variance explained by the first
principal component in each country when the seven indices from the RSF and
\text{V-Dem} datasets are combined. In other words, the PCA's linear
combinations refer to within-country annual variances. As can be seen, there is
only a single outlying country, and the performance is extremely good; almost
all variance is captured by the first PC. The second plot is similar except that
the period observed is restricted to the one allowed by the Freedom House
dataset. Although there is more variance, the performance is again
marvelous. The third and final plot shows the computation for all three
datasets, using the shorter period. Again: Almost 97\% is explained by the first
principal component. Thus, all of the indices can be concluded to measure the
same underlying semi-narrow definition for press freedom when the focus is on
within-country variances. To some extent, the conclusion also restates the
earlier points about the lack of variance exhibited by many of the indices.

The results are similar when a corresponding computation is carried out across
years---such that the PCA explains the linear combinations of the values across
all countries in a given year. The computation is summarized in Fig.~\ref{fig:
  pca countries}, which again shows separate estimates for the two periods. On
average, the first principal components explain as much as 98\% and 99\% of the
variances in the two periods, respectively. These shares are again good in the
context of comparative empirical research. The scores from the first PCs could
well be used in further computations, including regression
analysis. Alternatively, any single index might do as well---a point that points
toward the interchangeability assessment and the final regression results.

The multi-level regression results are summarized in Tables~\ref{tab: reg 1} and
\ref{tab: reg 2}. There are six brief points to make from the results. First,
the results are \change{fairly} similar in both tables; the estimates converge
between the full period and the shorter period between 2013 and 2020. Second,
all coefficients except the ones for the RSF dataset are positive, but this
result is a superficial artifact from the index's scale; as was noted, higher
values imply less freedom in the RSF dataset. \change{Also the coefficients for
  the V-Dem variables are positive in their sign. This reflects the scale of the
  variables; higher values imply more freedom. Thus, for instance, one unit
  increase in the censorship variable---which means there is less
  censorship---tends to increase deliberative democracy by $0.096$ according to
  Table~\ref{tab: reg 1}.} Third, the magnitudes are noteworthy because the
deliberative democracy index operates with a $(0, 1)$ scale. Fourth, the
standard deviation of the random country effects is much larger than the
standard deviation of the random annual effects. Fifth, the RSF index attains
the lowest BICs \change{in Table~\ref{tab: reg 1}}. This observation partially
reflects the already noted variance in the index when compared to the other
seven indices. Last but not least, all of the press freedom indices attain
coefficients with comparable magnitudes. To conclude: Both convergence validity
and interchangeability seem to hold.

\begin{table}[t!]
\centering
\caption{Regression Results: 2001 -- 2020 ($n = 2920$)}
\label{tab: reg 1}
\begin{tabular}{lrrrrrrrr}
\toprule
& \multicolumn{8}{c}{Model} \\
\cmidrule{2-9}
& 1. & 2. & 3. & 4. & 5. & 6. & 7. & 8. \\
\hline
Intercept & $0.422$ & -- & $0.296$ & $0.291$ & $0.211$ & $0.255$ & $0.262$ & $0.237$ \\
$\ln(\textmd{RSF} + 1)$ & $-0.034$ & -- & -- & -- & -- & -- & -- & -- \\
Freedom House & -- & -- & -- & -- & -- & -- & -- & -- \\
Censorship & -- & -- & $0.096$ & -- & -- & -- & -- & -- \\
Harassment  & -- & -- & -- & $0.116$ & -- & -- & -- & -- \\
Self-censorship & -- & -- & -- & -- & $0.100$ & -- & -- & -- \\
Media bias & -- & -- & -- & -- & -- & $0.093$ & -- & -- \\
Perspectives & -- & -- & -- & -- & -- & -- & $0.100$ & -- \\
Critical journalism & -- & -- & -- & -- & -- & -- & -- & $0.100$ \\
$\ln(\textmd{GDP per capita})$ & $0.015$ & -- & $0.008$ & $0.007$ & $0.018$ & $0.012$ & $0.009$ & $0.012$ \\
\hline
Std.~dev.~years & $0.010$ & -- & $0.004$ & $0.004$ & $<0.001$ & $0.006$ & $0.004$ & $0.007$ \\
Std.~dev.~countries & $0.217$ & -- & $0.128$ & $0.127$ & $0.142$ & $0.156$ & $0.160$ & $0.144$ \\
BIC & $-6708$ & -- & $-8687$ & $-8437$ & $-8200$ & $-8073$ & $-8204$ & $-8486$ \\
\bottomrule
\end{tabular}
%
\vspace{25pt}
%
\centering
\caption{Regression Results: 2013 -- 2020 ($n = 1168$)}
\label{tab: reg 2}
\begin{tabular}{lrrrrrrrr}
\toprule
& \multicolumn{8}{c}{Model} \\
\cmidrule{2-9}
& 1. & 2. & 3. & 4. & 5. & 6. & 7. & 8. \\
\hline
Intercept & $0.970$ & $0.010$ & $0.091$ & $0.099$ & $0.016$ & $0.019$ & $-0.042$ & $0.019$ \\
$\ln(\textmd{RSF} + 1)$ & $-0.231$ & -- & -- & -- & -- & -- & -- & -- \\
Freedom House & -- & $0.057$ & -- & -- & -- & -- & -- & -- \\
Censorship & -- & -- & $0.089$ & -- & -- & -- & -- & -- \\
Harassment  & -- & -- & -- & $0.087$ & -- & -- & -- & -- \\
Self-censorship & -- & -- & -- & -- & $0.076$ & -- & -- & -- \\
Media bias & -- & -- & -- & -- & -- & $0.067$ & -- & -- \\
Perspectives & -- & -- & -- & -- & -- & -- & $0.076$ & -- \\
Critical journalism & -- & -- & -- & -- & -- & -- &  -- & $0.091$ \\
$\ln(\textmd{GDP per capita})$ & $0.030$ & $0.035$ & $0.032$ & $0.031$ & $0.042$ & $0.041$ & $0.046$ & $0.038$ \\
\hline
Std.~dev.~years & $0.007$ & $0.008$ & $0.005$ & $0.005$ & $0.006$ & $0.007$ & $0.005$ & $0.008$ \\
Std.~dev.~countries & $0.149$ & $0.164$ & $0.122$ & $0.136$ & $0.147$ & $0.158$ & $0.146$ & $0.136$ \\
BIC & $-3061$ & $-3000$ & $-3605$ & $-3352$ & $-3206$ & $-3092$ & $-3176$ & $-3326$ \\
\bottomrule
\end{tabular}
\end{table}

\section{Conclusion}

This paper evaluated eight measures for press freedom across $146$ countries in
a period between 2001 and 2020. Based on principal component analysis,
convergence validity can be concluded to be good; to a reasonable degree, the
eight indices all measure the same semi-narrow definition for press
freedom. Based on regression analysis, also interchangeability holds. All in
all, the results replicate the earlier study of
\citet{Becker07}. \change{Despite these results, there remains methodological
  issues with the three datasets evaluated. In purely statistical terms, data
  from the RSF would be preferable as it provides more variation, but a major
  limitation with the RSF is that the organization has continuously changed its
  methodology for evaluating press freedom. Freedom House has been much more
  consistent in this regard. From this methodological viewpoint, it can be
  argued that the V-Dem project provides the best available source for academic
  research; the methodology is rigorous, all data is properly archived, and the
  evaluations are carried out by independent academics. A further benefit of the
  V-Dem project is that it provides separate indices for press freedom instead
  of one composite index. An alternative approach was suggested by
  \citet{Becker07}: It would be possible to combine the indices in order to
  increase the internal consistency of measurements.}

Some \change{further} limitations should be acknowledged. The conventional
concerns about comparative cross-country research apply. Among these concerns is
sampling; the datasets used were not random samples from a population of all
countries in the world. Although the coverage of $146$ countries is fairly
good---given that there were around two hundred countries in the world in 2020,
generalizability of the results is a good topic for further research. Also
interchangeability can be interpreted via generalizability; any given sample is
interchangeable with another same-sized one when the results remain identical or
at least highly similar~\citep{Kreiter20}. Another option would be to examine
different clusters or regimes of press freedom. A~more fundamental problem
relates to the noted national perceptions and cultural differences. These are a
problem in comparative communication research in particular because
communication is almost always dependent on given cultural
settings~\citep{Livingstone03}. However, the semi-narrow definition for press
freedom somewhat downplays this criticism; there is presumably a consensus among
journalists worldwide that issues like censorship, harassment, and violence
violate press freedom.

\change{Nevertheless, debates continue over the definitions for press
  freedom. Some have preferred even a narrower definition for press freedom,
  arguing that the right to criticize governments and hold them accountable
  forms the core of press freedom~\citep{WhittenWoodring17}. Such a narrow
  definition reflects a libertarian viewpoint to press freedom, which has been
  prevalent in the United States and other English-speaking countries. The media
  systems and also the question about press freedom continue to be different in
  other Western countries~\citep{Ruohonen21ICSET}. In this regard, all three
  datasets evaluated tend to share a similar semi-narrow definition for press
  freedom, which is understood mainly through interference by governments and
  other actors. As was noted, a broader definition would incorporate also
  aspects related to social responsibility, mediation of societal conflicts,
  public broadcasting, and other factors. Such a broader understanding about
  media and press freedom has been common in the Nordic countries. The absence
  of such factors is a limitation in all three datasets.}

\change{A further major issue is that the indices evaluated do not} consider the
freedom of digital press and digital journalism. \change{Although the V-Dem
  project provides separate indices for online media, these are not evaluated in
  terms of press freedom.} While the traditional threats are present for both
offline and online journalism, the digitalization of press has also brought new
threats---or at least old threats in new clothes. Allegedly, the harassment of
journalists and other acidity towards press happen increasingly in online
environments, including social media in particular. To some extent, press is a
part of the problem rather than the solution. A free press is not necessary a
trustworthy press~\citep{Soon17}. Increasingly, press in many countries is
entangled in the same web of engagement and emotionality, clicks and metrics,
and audience-building at all costs, not to mention the so-called culture war
that keeps the show running for many journalists and audiences
alike~\citep{Ackerley21}. \change{Social media and technology companies have
  also significantly worsened the financial situation of media throughout the
  world. The revenues from advertising increasingly go to technology companies
  instead of media and publishers. The recommender systems of these companies
  nowadays often dictate what news people consume. Consequently, technological
  factors have presumably contributed to the biases and plurality of media, but
  none of the three datasets evaluated consider such factors.} These issues
point toward the much wider and more pressing debate about the freedom of speech
in online environments.

\section{Funding}

This paper was partially funded by the Strategic Research Council at the Academy of Finland (grant number~327391).

\section{Conflicts of Interest}

There is a conflict of interest with researcher funded by the grant no.~327391 from the Strategic Research Council at the Academy of Finland.


\end{document}